\documentclass[preprintnumbers,showpacs,superscriptaddress,nofootinbib]{revtex4}
\usepackage{amsmath,amsfonts,amssymb}
\usepackage{graphicx}
\usepackage{epsfig}
\usepackage{color}
\usepackage[dvipsnames,svgnames,x11names]{xcolor}
\usepackage{slashed} 
\usepackage{url}
\usepackage{subfigure}
\usepackage{multirow} 

\def\GeV{\mathrm{GeV}} 

\makeatother

\allowdisplaybreaks 

\begin{document}

\title{ Testing the electroweak phase transition and electroweak  baryogenesis at LHC and a circular electron-positron collider }

\author{Fa Peng Huang\footnote{Electronic
address: huangfp@ihep.ac.cn}}
\affiliation{Theoretical Physics Division, Institute of High Energy Physics,
Chinese Academy of Sciences, P.O.Box 918-4, Beijing 100049, P.R.China}


\author{Pei-Hong Gu}
\affiliation{Department of Physics and Astronomy, Shanghai Jiao Tong University,
800 Dongchuan Road, Shanghai 200240, China}

\author{Peng-Fei Yin}
\affiliation{Key Laboratory of Particle Astrophysics, Institute of High Energy Physics,
Chinese Academy of Science, P.O.Box 918-3, Beijing 100049, P. R. China}

\author{Zhao-Huan Yu}
\affiliation{Key Laboratory of Particle Astrophysics, Institute of High Energy Physics,
Chinese Academy of Science, P.O.Box 918-3, Beijing 100049, P. R. China}
\affiliation{ARC Centre of Excellence for Particle Physics at the Terascale,
School of Physics, The University of Melbourne, Victoria 3010, Australia}

\author{Xinmin Zhang}
\affiliation{Theoretical Physics Division, Institute of High Energy Physics,
Chinese Academy of Sciences, P.O.Box 918-4, Beijing 100049, P.R.China}

\begin{abstract}
We study the collider phenomenology of the electroweak phase transition and
electroweak baryogenesis in the framework of the effective field theory. Our study shows that
the effective theory using the dimension-6 operators can enforce strong first order phase transition and provide sizable CP violation to realize a successful electroweak baryogenesis. Such dimension-6 operators can induce interesting Higgs phenomenology that can be verified at colliders such as the LHC and the planning CEPC. We then demonstrate that this effective theory can originate from vector-like quarks and the triplet Higgs.

\end{abstract}


\maketitle

\section{introduction}
A longstanding problem in particle physics and cosmology is to unravel the origin of baryon asymmetry of the universe (BAU), which is
quantified by the baryon-to-photon ratio $\eta=n_B/n_{\gamma}=6.05(7)\times 10 ^{-10}$~\cite{Ade:2013zuv,Agashe:2014kda}.
The current value $\eta$ is obtained from investigations of the big bang nucleosynthesis or the power spectrum of the cosmic microwave background radiation.
To generate the BAU (baryogenesis), it needs three necessary conditions, i.e. baryon number violation, C and CP violation, and departure from equilibrium dynamics or CPT violation~\cite{Sakharov:1967dj}.
To satisfy the three conditions for baryogenesis,  many possible  mechanisms, such as  Planck-scale baryogenesis, GUT baryogenesis,
Affleck-Dine baryogenesis, leptogenesis, and electroweak (EW) baryogenesis have been proposed \cite{Dine:2003ax}, but after
the discovery of the $125$ GeV Higgs boson \cite{Aad:2012tfa,Chatrchyan:2012xdj} at the
LHC, EW baryogenesis \cite{Kuzmin:1985mm,Trodden:1998ym} becomes a popular and testable scenario for explaining the BAU \cite{Morrissey:2012db}, wherein the BAU
is driven by the EW sphaleron (baryon number violation) and the generation of CP asymmetry (C and CP violation) at the time of the EW phase transition (departure from equilibrium dynamics). For efficient production of BAU, a strong first order  phase transition (SFOPT) and sizable CP violation are necessary.
However,  in the standard model (SM), the $125$ GeV Higgs boson is too heavy for a  SFOPT \cite{Morrissey:2012db}, and the CP violation from CKM matrix is too weak.

Another urgent problem in particle physics after the discovery of the Higgs boson
is to explore the true shape of the Higgs potential, the nature of the EW spontaneous symmetry breaking,  and the type of the EW phase transition.
However, the current experiments at the LHC only provide  us with a rough picture of these problems.
For the Higgs potential, we know nothing but the quadratic oscillation around the vacuum expectation value ({\rm vev}) $v$ with the $125~\rm GeV$ mass.
Understanding these problems can also help to understand the above EW baryogenesis problem.

To provide both the SFOPT and  sizable CP  violation for EW baryogenesis,
we follow the effective field theory (EFT) approach in Refs.~\cite{Zhang:1992fs,Zhang:1993vh,Whisnant:1994fh,Zhang:1994fb,Huang:2015bta})
to realize the EW baryogenesis
by introducing dimension-6 operators $-x_{u}^{ij}\frac{\phi^\dagger_{}\phi}{\Lambda^2_{}}\bar{q}^{}_{Li}\tilde{\phi} u^{}_{Rj}+\textrm{H.c.}-\frac{\kappa}{\Lambda^2_{}}(\phi^\dagger_{}\phi)^3$ in this paper.
The dimension-6 operators will modify the shape of the Higgs  potential and yield distinctive signals at the LHC, such as the different Higgs pair production behavior.
Due to experimental precision of the LHC, LHC may only give some hints and be difficult to precisely
test this type of EW baryogenesis scenario.
However, the Circular Electron-Positron Collider (CEPC)\cite{CEPC-SPPCStudyGroup:2015csa} has the ability to precisely test this type of the EW  phase transition and
the EW baryogenesis, which is the main goal of the CEPC. Detailed discussions on testing the  EW phase transition
and the EW baryogenesis will be given in the following. For completeness, a renormalizable extension model is
given to obtain the needed dimension-6 operators for the EW baryogenesis.

In Sec.~\ref{sec:Model}, we describe the effective operators in the EFT framework,
and show that the dimension-6 operators can change the Higgs potential, realize the SFOPT, and provide the
CP violation source for the EW baryogenesis.
In Sec.~\ref{sec:collider}, we discuss the collider phenomenology of this EW baryogenesis scenario induced by the dimension-6
operators at the LHC and the future CEPC, especially the Higgs phenomenology.
In Sec.~\ref{sec:ren},  the possible renormalizable model is given.
Finally, we conclude in Sec.~\ref{sec:sum}.

\section{An effective theory for EW baryogenesis}\label{sec:Model}
In this paper, the EFT approach is adopted
to provide the needed SFOPT and CP violation for the EW baryogenesis.
This allows us
to derive model-independent predictions and constraints on the EW baryogenesis using the following effective
operators:
\begin{equation}
\mathcal{L}_\mathrm{eff}=\mathcal{L}_\mathrm{SM}+\Sigma_i \frac{1}{\Lambda^{d_i-4}} \kappa_i \mathcal{O}_i,
\end{equation}
where $d_i$ are the dimensions of the new operators, and $\mathcal{O}_i$
are invariant under the SM gauge
symmetry and contain only the SM fields. The parameters
$\kappa_i$ is the Wilson coefficient, which can be determined by matching the full
theory to the effective operators. $\Lambda$ is the cutoff energy scale, at which the effective theory breaks down.

\subsection{The effective Lagrangian}
In addition to the relevant Lagrangian of the Higgs sector in the SM
\begin{eqnarray}
\label{sm}
\mathcal{L}_{\textrm{SM}}^{}&\supset&-y_{u}^{ij}\bar{q}^{}_{Li}\tilde{\phi} u^{}_{Rj}+\textrm{H.c.}-\mu^2_{}\phi^\dagger_{}\phi-\lambda(\phi^\dagger_{}\phi)^2_{}\,,
\end{eqnarray}
we commence our study on the  EW baryogenesis from the effective Lagrangian with the following dimension-6 operators~\cite{Zhang:1992fs,Zhang:1993vh,Whisnant:1994fh,Zhang:1994fb,Huang:2015bta}:
\begin{eqnarray}
\label{dim6}
\delta\mathcal{L}&=&-x_{u}^{ij}\frac{\phi^\dagger_{}\phi}{\Lambda^2_{}}\bar{q}^{}_{Li}\tilde{\phi} u^{}_{Rj}+\textrm{H.c.}-\frac{\kappa}{\Lambda^2_{}}(\phi^\dagger_{}\phi)^3_{}\,,
\end{eqnarray}
where $\phi$ is the Higgs doublet field and  $\tilde{\phi}=i \tau_2 \phi^*$.  $q_L^{}$ and $u_R^{}$ are the left-handed quarks and the right-handed up-type quarks, respectively. Thus,
\begin{eqnarray}
\!\!\!\!\!\!&&\!\!\begin{array}{l}\phi(1,2,-\frac{1}{2})\end{array}\!\!=\!\!\left[\!\begin{array}{l}\phi^{0}_{}\\
[1mm]
\phi^{-}_{}\end{array}\!\!\right]\!,
~\begin{array}{l}q^{}_L(3,2,+\frac{1}{6})\end{array}\!\!=\!\!\left[\!\begin{array}{l}u^{}_L\\
[1mm]
d^{}_L\end{array}\!\!\right]\!,~\begin{array}{l}u^{}_{R}(3,1,+\frac{2}{3})\end{array}\!.\nonumber\\
\!\!\!\!\!\!&&
\end{eqnarray}
Here and hereafter the brackets following the fields describe the transformations under the $SU(3)^{}_c\times SU(2)^{}_L\times U(1)^{}_{Y}$ gauge group.
The effective Lagrangian in Eq.~(\ref{dim6}) can be obtained in many models beyond the SM, such as some strong dynamics models.
In Sec.~\ref{sec:ren}, a renormalizable model is given.
The early study of the SFOPT using the  $(\phi^\dagger_{}\phi)^3$  effective opertor was first given in Ref.~\cite{Zhang:1992fs},
and the recent studies can be found in Refs.~\cite{Grojean:2004xa, Delaunay:2007wb, Grinstein:2008qi,Chung:2012vg,Ham:2004zs,Bodeker:2004ws,Chu:2015nha,Huang:2016odd}.

\subsection{SFOPT and the true Higgs potential}

\begin{figure}
  \centering
  \includegraphics[width=.4\textwidth]{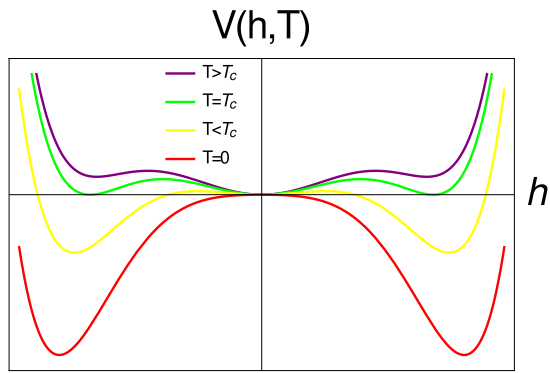}\\
  \caption{The schematic diagram of the Higgs potential for SFOPT at different temperatures.}\label{vh6}
\end{figure}

From  Eqs.~(\ref{sm}) and  (\ref{dim6}), the  potential of the Higgs boson field can be written as
\begin{equation}\label{v0}
V(\phi)=\mu^2 \phi^{\dag}\phi + \lambda (\phi^{\dag}\phi)^2+ \frac{\kappa(\phi^{\dag}\phi)^3}{\Lambda^2}.
\end{equation}
By introducing the  non-renormalizable term $(\phi^{\dagger} \phi)^3$ in the potential, the SFOPT may be realized.
When the SFOPT is considered, we
can simplify the potential by substituting $\phi$ with $h/\sqrt{2}$:
\begin{equation}\label{v0}
V_{\rm tree}(h) = \frac{1}{2}\mu^2 h^2 + \frac{\lambda}{4} h^4 + \frac{\kappa}{8\Lambda^2} h^6.
\end{equation}
Using the methods in Refs.~\cite{Quiros:1999jp,Dolan:1973qd}, the full finite-temperature
effective potential up to one-loop level is composed of three parts,
\begin{equation}\label{fullpotential}
 V_\mathrm{eff}(h,T)=V_\text{tree}(h)+
V_1^{T=0}(h)+\Delta V_1^{T\neq 0}(h,T),
\end{equation}
where $V_\text{tree}(h)$ is the tree-level potential, $V_1^{T=0}(h)$ is the Coleman-Weinberg
potential at zero temperature, and $\Delta V_1^{T\neq 0}(h)$ is the leading thermal correction
including the daisy resummation.
However, in this scenario, the dominant contribution for the SFOPT comes from the tree level barrier as shown
in Fig.~\ref{vh6}, and thus the  effective potential with finite temperature effects can be approximated as
\begin{equation}\label{Veff}
	V_{\rm eff}(h, T) \approx \frac{1}{2} \left( \mu^2 + c \, T^2 \right) h^2 + \frac{\lambda}{4} h^4 + \frac{\kappa}{8\Lambda^2} h^6.
\end{equation}
The  finite temperature effects are included in the coefficient $c$ of the thermal mass with
\begin{equation}
c=\frac{1}{16}(g'^2+3 g^2+4 y_t^2+4\frac{m_h^2}{v^2}-12\frac{\kappa v^2}{\Lambda^2}),
\end{equation}
where $g$ and $g'$ are the $SU(2)_L$ and $U(1)_Y$ gauge couplings, and $y_t$ is the top quark Yukawa coupling in the SM.
To keep the observed Higgs boson mass $m_h=125~\rm GeV$ and the {\rm vev}  $v$ , $\lambda$ and  $\mu^2$  should  be changed as
\begin{eqnarray}
	\lambda
	&=& \frac{m_h^2}{2 v^2} \left( 1 - \frac{\Lambda_{\rm max}^2}{\Lambda^2} \right)  \label{lambda}, \\
	\mu^2
	&=& \frac{m_h^2}{2} \left( -1+\frac{\Lambda_{\rm max}^2}{2 \Lambda^2} \right) \label{musq},
\end{eqnarray}
with $\tilde\Lambda_{\rm max} \equiv \sqrt{3\kappa} v^2 / m_h$.
In the limit $\Lambda \to \infty$, $\lambda$ and  $\mu^2$ are recovered to the SM cases as
\begin{eqnarray}
	\lambda(\Lambda \to \infty)
	&=& \frac{m_h^2}{2 v^2}  \label{smlambda}, \\
	\mu^2(\Lambda \to \infty)
	&=& -\frac{m_h^2}{2}  \label{smmusq}.
\end{eqnarray}
The Higgs potential in Eq.~(\ref{v0}) could trigger the desired
spontaneous symmetry breaking as in the SM.
To realize the SFOPT, it needs $\mu^2 + c \, T^2 > 0$  to make the EW-symmetric vacuum stable, $\lambda < 0$  to reverse the potential, and the $h^6$ term to stabilize the EW-broken vacuum as shown in Fig.~\ref{vh6}.
In order to guarantee the SFOPT,  $\lambda$ must be negative under the above approximation, and thus, from Eq.~(\ref{lambda}),
we obtain an upper bound on $\Lambda$, namely $\Lambda < \tilde\Lambda_{\rm max}\approx 840\sqrt{\kappa}~\GeV$.
From the requirements of perturbativity, $\kappa < 4 \pi$. If we choose a larger
$\kappa$, we can get a larger upper bound $\tilde\Lambda_{\rm max}$. For example,
if $\kappa=12.5$, then $\tilde\Lambda_{\rm max}=3~\rm TeV$. This large upper bound
leaves large space to discuss the phenomenology at the $14$ TeV LHC.

\begin{figure}
  \centering
  \includegraphics[width=.35\textwidth]{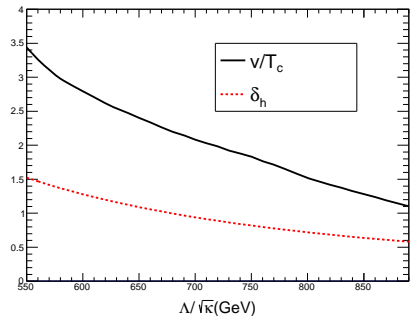}\\
  \caption{The dependence of the washout parameter and the modification of the trilinear Higgs boson coupling on $\Lambda/\sqrt{\kappa}$.}\label{wash}
\end{figure}

Using the standard techniques in  the study of the EW baryogenesis, we obtain the phase transition critical temperature $T_c$ and the washout parameter $v(T_c)/T_c$ as the following:
\begin{eqnarray}
	T_c & =& \frac{\sqrt{\lambda^2 \Lambda^2- 4 \kappa \mu^2}}{2\sqrt{c \kappa}} ,	 \label{Tc} \\
	\frac{v(T_c)}{T_c} & =& \frac{2\Lambda \sqrt{-c\lambda}}{\sqrt{\lambda^2 \Lambda^2- 4 \kappa \mu^2}} \, . \label{vTcoverTc}
\end{eqnarray}
$T_c>0$ gives a lower bound $\tilde\Lambda_{\rm min} \equiv \tilde\Lambda_{\rm max} / \sqrt{3} =\sqrt{\kappa} v^2 / m_h$.
For $m_h = 125~\GeV$, this gives $\tilde\Lambda_{\rm min} \approx 480\sqrt{\kappa} \rm~GeV$.

After including the full one-loop results as given in Refs.~\cite{Delaunay:2007wb,Bodeker:2004ws},
the dependence of the washout parameter on $\Lambda/\sqrt{\kappa}$ is shown in Fig.~\ref{wash},
and it shows that the washout condition for the SFOPT,
\begin{equation}
\frac{v(T_c)}{T_c}   \gtrsim 1,
\end{equation}
can be easily satisfied roughly from $550$ to $890$ GeV. For instance, taking  $\Lambda/\sqrt{\kappa}=620$ GeV, the washout parameter
$v/T_c=2.6$ for the full one-loop case
and $v/T_c=2.8$ for the leading thermal approximation case.   It is worth noticing that for $\Lambda/\sqrt{\kappa} \lesssim 550$ GeV, the phase transition will
not occur by the ``metastability"~\cite{Bodeker:2004ws} and for $\Lambda/\sqrt{\kappa}  > 840$ GeV, the parameter $\lambda$ becomes positive.
Thus, after considering the full one-loop results and the metastability, the allowed range changes from the approximation result
$480~\rm GeV<\Lambda/\sqrt{\kappa}<840$ GeV
to roughly $550~\rm GeV \lesssim \Lambda/\sqrt{\kappa} \lesssim 890$ GeV.

An interesting consequence is that the requirement of the SFOPT would lead to obvious modification of the trilinear Higgs boson
coupling as
\begin{equation}
\mathcal{L}_{hhh}= -\frac{1}{3!} (1+ \delta_h) A_{h} h^3,
\end{equation}
where $A_{h}=3 m_h^2/v$ is the trilinear Higgs boson coupling in the SM,
and $\delta_h$ is the modification of the trilinear Higgs coupling induced by the dimension-6 operator.
In this scenario, $\delta_h$ roughly varies from $0.6$ to $1.5$ in the allowed parameter space as
shown in Fig.~\ref{wash}.
Hints of the  large deviation
of the trilinear Higgs coupling may be seen  at the 14 TeV LHC, and its precise probe
may be obtained in the future CEPC~\cite{Noble:2007kk,Katz:2014bha,Curtin:2014jma}, which will be discussed in details below.

\subsection{CP violation source}
Besides the SFOPT,
the BAU also needs sizable source of CP violation, which can be provided by
the dimension-6 operator $-x_{u}^{ij}\frac{\phi^\dagger_{}\phi}{\Lambda^2_{}}\bar{q}^{}_{Li}\tilde{\phi} u^{}_{Rj}+\textrm{H.c.}$  in Eq.~(\ref{dim6}).
For simplicity,  only the top quark is considered in this paper.
Therefore, this effective operator
can induce the anomalous top quark Yukawa  coupling
$\frac{v^2}{2 \Lambda^2} h[\mathrm{Re}(x_{u}^{ij})\bar{t}t+i\,\mathrm{Im}(x_{u}^{ij})\bar{t}\gamma^5 t]$.
For simplicity, we denote
$\mathrm{Re}(x_{u}^{33})=a$ and $\mathrm{Im}(x_{u}^{33})=b$. Then the CP violation source from  Eq.~(\ref{dim6}) can
be rewritten as
\begin{equation}
\frac{v^2}{2 \Lambda^2} h(a \bar{t}t+i b \bar{t}\gamma^5 t).
\end{equation}
To avoid the non-minimal flavour violation~\cite{Lebedev:2010zg,Huber:2006ri}, we assume that  $x_{u}^{ij}=0$ except for $x_{u}^{33}\neq 0$.
Hence, the Lagrangian for  the effective interaction between the top quark and the Higgs boson  is parameterized as
\begin{equation}
\mathcal{L}=-\frac{m_t}{v} h \bar{t} (1+\delta^+_t+i\delta^-_t \gamma^5)t,
\end{equation}
with
\begin{equation}\label{tplus}
\delta_t^{+}=\frac{a v^3}{2 \Lambda^2 m_t},
\end{equation}
and
\begin{equation}\label{tplus}
\delta_t^{-}=\frac{b v^3}{2 \Lambda^2 m_t}.
\end{equation}
The $i\gamma^5$ term is CP-odd, and would provide the CP violation source for the EW baryogenesis.
And thus, the top quark  acquires a
complex mass  inside the
bubble walls, which is expressed  as~\cite{Espinosa:2011eu,Cline:2012hg}
\begin{equation}
	m_t(z) = \frac{m_t}{v} (1+\delta^+_t+i\delta^-_t \gamma^5)h(z)\equiv
		|m_t(z)| e^{i\Theta(z)}.
\end{equation}
Here, $z$ is just the coordinate transverse to the bubble wall, and
$\Theta$ is the CP violation phase.
Using the approximate method
with the more recent and complete transport equations in Refs.~\cite{Cline:2012hg,Fromme:2006wx,Cline:2011mm},
the baryon asymmetry is given by\footnote{The expressions of the baryon asymmetry in Refs.~\cite{Cline:2012hg,Cline:2011mm} are based  on
the more recent result in Ref.~\cite{Fromme:2006wx}, which improves the result in  Ref.~\cite{Bodeker:2004ws} with the more recent and complete transport equations. The baryon asymmetry expression in Ref.~\cite{Fromme:2006wx} is obtained in the wall frame. In this work, we focus on the study of the collider phenomenology rather than the transport details, which are given in Refs.~\cite{Bodeker:2004ws,Fromme:2006wx}.}
\begin{equation}
	\eta_B = {405\Gamma_{\rm sph}\over 4\pi^2v_\mathrm{wall} g_*T}\int dz\, \mu_{B_L}
	f_{\rm sph}\,e^{-45\, \Gamma_{\rm sph}|z|/(4 v_\mathrm{wall})},
\label{baueq}
\end{equation}
which depends on the sphaleron washout
parameter $v_c/T_c$, the bubble wall velocity $v_\mathrm{wall}$,  the
bubble wall thickness $L_\mathrm{wall}$, and so on.
Here, $f_{\rm sph} \approx \rm min(1, 2.4\rm T/\Gamma_{\rm sph} e^{-40v/T})$.
Preliminary numerical estimation gives a rough estimation on the anomalous top quark Yukawa coupling as
\begin{eqnarray}
\label{cp}
\delta^-_t= \mathcal{O}(0.01-1)\,,
\end{eqnarray}
which can  provide  sizable CP violation source for a successful EW baryogenesis.
Due to the fact that the exact calculations of the baryon asymmetry $\eta_B$ rely on the improvements of the non-perturbative dynamics,
such as the bubble wall dynamics, we can only obtain a rough constraint on the anomalous top quark Yukawa coupling
from $\eta_B$  and
we will discuss how to  constrain the  anomalous coupling from experiments in particle physics,
where more accurate constraints might be obtained.

\section{collider phenomenology from the LHC to the CEPC}\label{sec:collider}
From the above discussions on the EW baryogenesis induced by the dimension-6 operators in Eq.~(\ref{dim6}),
we see that the realization of the EW baryogenesis will lead to obvious  modifications of  the trilinear Higgs boson coupling
and the top quark Yukawa coupling  on the SM values, which
may leave hints at the 14 TeV LHC, and further be precisely tested at the future CEPC.
The deviations due to Eq.~(\ref{dim6}) can be written as
 \begin{equation}
\label{lag}
-{\cal L}=\frac{1}{3!}\left(\frac{3m_h^2}{v}\right)\,(1+ \delta_h)\,h^3
\ + \
\frac{m_t}{v} h \bar{t} (1+\delta^+_t+i\delta^-_t \gamma^5)t\,.
\end{equation}
In the SM, $\delta_h=\delta^+_t=\delta^-_t=0$.

\subsection{Collider phenomenology at the LHC}
Now, we discuss the constraints and possible signals of the anomalous couplings $\delta_h$, $\delta^+_t$, and $\delta^-_t$ at the LHC.
Firstly, we consider the phenomenology  of the anomalous top quark Yukawa coupling.

In low energy experiments, the CP-odd top Yukawa coupling is constrained from the electron electric dipole moments (EDM), neutron EDM, and Hg EDM, etc taking into account all uncertainties in the low energy matrix elements\cite{Zhang:1994fb,Shu:2013uua,Brod:2013cka,Inoue:2014nva,Chien:2015xha,Baron:2013eja}.
Among these low energy experiments, the electron EDM give the strong constraints
$\delta_t^- < \mathcal{O}(0.01)$, but depends on the so far not measured SM electron Yukawa coupling.
Introducing new baryogenesis unrelated CP violation source~\cite{Bian:2014zka,Fuyuto:2015ida} or modifying the electron Yukawa coupling can
relax the stringent constraints from the electron EDM.
The neutron EDM does not suffer from the electron Yukawa coupling (it can be induced purely from the interaction of the top quark and Higgs boson)
and its bound, including all uncertainties, roughly gives
$\delta_t^- < \mathcal{O}(0.1)$, which still leaves room for successful baryogenesis.
The Hg EDM gives no constraints anymore after the uncertainties are taken into account.
Detailed discussion on the cancellations between the EDM with a CP violation Higgs sector
can be found in Ref.~\cite{Huber:2006ri}. In the following discussions, we assume that
the constraints from the EDMs are relaxed and investigate the collider constraints.

The modifications of the anomalous top quark Yukawa coupling would also affect the single Higgs production $gg \rightarrow h$
and the Higgs pair production $gg \rightarrow hh$ via gluon fusions and the Higgs decay to photons $h\rightarrow \gamma\gamma$
at hadron colliders. It is possible to constrain the deviations to be of $ \mathcal{O}(0.1)$  at the LHC \cite{Nishiwaki:2013cma}.
The anomalous top quark Yukawa  coupling would modify the Higgs couplings to $gg$ and $\gamma\gamma$.
The loop induced Higgs couplings $g_{hgg}$ and $g_{h\gamma \gamma}$ can be parameterized as~\cite{Nishiwaki:2013cma,Khatibi:2014bsa,Liu:2014rba}
\begin{eqnarray}
g_{hgg}^2/g_{hgg,\mathrm{SM}}^2 &\simeq& (1+\delta_t^+)^2+ 0.11 \delta_t^+ (1+\delta_t^+)  +2.6 (\delta_t^-)^2 , \label{hloopgg}
\\
g_{h\gamma\gamma}^2/g_{h\gamma\gamma,\mathrm{SM}}^2 &\simeq& (1-0.28\delta^+_t)^2+(0.43\delta_t^-)^2. \label{hloopgaga}
\end{eqnarray}
Unlike the associated Higgs production, both CP-even and CP-odd top quark Yukawa  couplings enter the Higgs decays to $gg$ and $\gamma\gamma$.
It can be seen that $\delta_t^-$ has important contribution to $g_{hgg}$ from Eq.~(\ref{hloopgg}).
The numerical constraints will be discussed in the subsection of CEPC below.

\begin{figure}
  \centering
  \includegraphics[width=.5\textwidth]{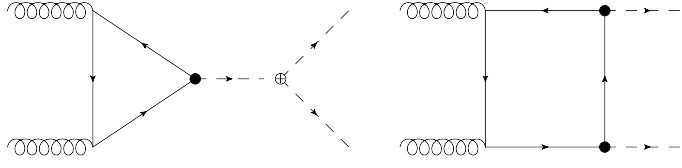}\\
  \caption{The sample leading order Feynman diagrams for the Higgs boson pair production induced by the top quark loop at the $14$ TeV LHC.}\label{gghh}
\end{figure}

Meanwhile, these three deviations will modify the behavior of the Higgs boson pair production at the LHC. The sample Feynman diagrams for the Higgs boson
pair production  are shown in Fig.~\ref{gghh}.
The kinematic invariants are defined as
$\hat{s}=(p_a+p_b)^2$,
$\hat{t}=(p_a+p_c)^2$, and
$\hat{u}=(p_b+p_c)^2$ with $p_a+p_b+p_c+p_d=0$.
And we also define the following kinematic variables:
\begin{displaymath}
\mathcal{S} = {\hat s}/m_t^2, \hspace{0.5 cm}
\mathcal{T} = {\hat t}/m_t^2, \hspace{0.5 cm}
\mathcal{U} = {\hat u}/m_t^2,  \hspace{0.5 cm}
\rho_c = m_h^2/m_t^2, \hspace{0.5 cm}
\mathcal{T}_1 =\mathcal{T} - \rho_c, \hspace{0.5 cm}
\mathcal{U}_1 =\mathcal{U} - \rho_c, \hspace{0.5 cm}
\mathcal{P}(\hat{s})= \frac{3m_h^2}{\hat{s}-m_h^2+im_h\Gamma_h}.
\end{displaymath}
Using \texttt{FeynCalc}~\cite{Mertig:1990an}, the partonic differential
cross section for the $g(p_a)g(p_b) \to h(p_c)h(p_d)$ process can be
obtained as
 \begin{eqnarray}
\frac{d\hat\sigma(gg\to hh)}{d\hat{t}}
& =  &
\frac{G_F^2 \alpha_s^2}{512(2\pi)^3} \biggr\{
\Big| (1+ \delta_h) (1+\delta^+_t)\mathcal{P}(\hat{s})F_\triangle^A
+(1+\delta^+_t)^2F_\Box^{AA}+(\delta^-_t)^2F_\Box^{BB} \Big|^2
\nonumber \\[2mm]
&+&
\Big| (1+\delta^+_t)\delta^-_tG_\Box^{AB} \Big|^2+
\Big| (1+\delta^+_t)^2G_\Box^{AA}+(\delta^-_t)^2G_\Box^{BB} \Big|^2
\nonumber \\[2mm]
&+&
\Big| (1+\delta_h)\delta^-_t\mathcal{P}(\hat{s})F_\triangle^B
+(1+\delta^+_t)\delta^-_tF_\Box^{AB} \Big|^2   \biggr \}\,, \label{psigma}
\end{eqnarray}
where $F_\triangle^A$, $F_\Box^{AA}$,  $F_\Box^{BB}$,  $F_\triangle^B$, $F_\Box^{AB}$, and $G_\Box^{AA}$
are the  corresponding form factors:
\begin{eqnarray}
F_\triangle^A & = & \frac{2}{\mathcal{S}} \left\{ 2+(4 - \mathcal{S}) m_t^2 C_{ab} \right\},
\end{eqnarray}
\begin{eqnarray}
F_\Box^{AA}&=&\frac{1}{\mathcal{S}^2}( 4 S + 8 \mathcal{S} m_t^2 C_{ab}
-2\mathcal{S}(\mathcal{S}+2 \rho_c-8) m_t^4(D_{abc}+D_{bac}+D_{acb})  \nonumber  \\
&+& (2 \rho_c-8) m_t^2 ( \mathcal{T}_1 C_{ac} + \mathcal{U}_1 C_{bc}
+ \mathcal{U}_1 C_{ad} + \mathcal{T}_1 C_{bd} - (\mathcal{T}\mathcal{U}-\rho_c^2) m_t^2 D_{acb}) ),
\end{eqnarray}
\begin{eqnarray}
G_\Box^{AA} &=& \frac{1}{\mathcal{S}(\mathcal{T}\mathcal{U}-\rho_c^2)}( (\mathcal{T}^2+\rho_c^2-8\mathcal{T})
m_t^2 ( \mathcal{S} C_{ab} + \mathcal{T}_1 C_{ac} +\mathcal{T}_1 C_{bd} - \mathcal{S} \mathcal{T} m_t^2 D_{bac} ) \nonumber \\
&+&(\mathcal{U}^2+\rho_c^2-8\mathcal{U})m_t^2( \mathcal{S} C_{ab} + \mathcal{U}_1 C_{bc}
+\mathcal{U}_1 C_{ad} - \mathcal{S}\mathcal{U} m_t^2 D_{abc} )  \nonumber \\
&-& (\mathcal{T}^2+\mathcal{U}^2-2\rho_c^2) (\mathcal{T}+\mathcal{U}-8) m_t^2 C_{cd}  \nonumber \\
&-& 2(\mathcal{T}+\mathcal{U}-8)(\mathcal{T}\mathcal{U}-\rho_c^2) m_t^4(D_{abc}+D_{bac}+D_{acb}) ),
\end{eqnarray}
\begin{equation}
F_\triangle^B =  -2 m_t^2 C_{ab},
\end{equation}
\begin{equation}
F_\Box^{AB} = -2 m_t^4 (D_{abc}+D_{bac}+D_{acb}),
\end{equation}
\begin{eqnarray}
G_\Box^{AB} &=& \frac{1}{\mathcal{S}(\mathcal{T}\mathcal{U}-\rho_c^2)}( (\mathcal{U}^2-\rho_c^2)
m_t^2 ( \mathcal{S} C_{ab} + \mathcal{U}_1 C_{bc} + \mathcal{U}_1 C_{ad} - \mathcal{S}\mathcal{U} m_t^2 D_{abc} ) \nonumber \\
&-&(\mathcal{T}^2-\rho_c^2) m_t^2
(  \mathcal{S} C_{ab} + \mathcal{T}_1 C_{ac} + \mathcal{T}_1 C_{bd} - \mathcal{S}\mathcal{T} m_t^2 D_{bac} ) \nonumber \\
&+&( (\mathcal{T}+\mathcal{U})^2-4\rho_c^2)(\mathcal{T}-\mathcal{U})m_t^2C_{cd}  \nonumber \\
&+& 2(\mathcal{T}-\mathcal{U})(\mathcal{T}\mathcal{U}-\rho_c^2) m_t^4(D_{abc}+D_{bac}+D_{acb}) ),
\end{eqnarray}
\begin{eqnarray}
F_\Box^{BB}&=&\frac{1}{\mathcal{S}^2}( 4 \mathcal{S} + 8 \mathcal{S} m_t^2 C_{ab}
-2\mathcal{S}(\mathcal{T}+\mathcal{U}) m_t^4(D_{abc}+D_{bac}+D_{acb})  \nonumber  \\
&+& 2 \rho_c m_t^2 ( \mathcal{T}_1 C_{ac} + \mathcal{U}_1 C_{bc}
+ \mathcal{U}_1 C_{ad} + \mathcal{T}_1 C_{bd} - (\mathcal{T}\mathcal{U}-\rho_c^2) m_t^2 D_{acb}) ),
\end{eqnarray}
\begin{eqnarray}
G_\Box^{BB} &=& \frac{1}{\mathcal{S}(\mathcal{T}\mathcal{U}-\rho_c^2)}( (\mathcal{T}^2+\rho_c^2)
m_t^2 ( \mathcal{S} C_{ab} + \mathcal{T}_1 C_{ac} +\mathcal{T}_1 C_{bd} - \mathcal{S}\mathcal{T} m_t^2 D_{bac} ) \nonumber \\
&+&(\mathcal{U}^2+\rho_c^2)m_t^2( \mathcal{S} C_{ab} + \mathcal{U}_1 C_{bc}
+\mathcal{U}_1 C_{ad} - \mathcal{S}\mathcal{U} m_t^2 D_{abc} ) \nonumber \\
&-& (\mathcal{T}^2+\mathcal{U}^2-2\rho_c^2) (\mathcal{T}+\mathcal{U}) m_t^2 C_{cd} \nonumber \\
&-& 2(\mathcal{T}+\mathcal{U})(\mathcal{T}\mathcal{U}-\rho_c^2) m_t^4(D_{abc}+D_{bac}+D_{acb}) ).
\end{eqnarray}
The definitions of the scalar integrals $C_{ij}$ and $D_{ijk}$ can be found in the appendix.
In the case of $\delta_t^+=\delta_t^-=0$,
there has been extensive study on extracting the $\delta_h$ from LHC data~\cite{Barger:2013jfa,Shao:2013bz,Papaefstathiou:2012qe,
Goertz:2013kp,Chen:2014xra,Li:2015yia,He:2015spf,Cao:2015oxx,Dawson:2015oha,Huang:2015tdv}.
If we further let $\delta_h=0$, the differential cross section here can reduce to the SM case as in Refs.~\cite{Glover:1987nx,Plehn:1996wb}.

\begin{figure}
  \centering
  \includegraphics[width=.4\textwidth]{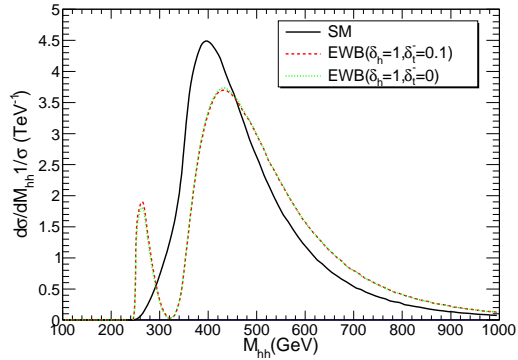}\\
  \caption{The normalized  invariant mass distribution of the Higgs boson pair for $PP \to hh$ at the $14$ TeV LHC.
  The solid black line represents the result for SM. The red dashed line represents the result for EWBG with $\delta_h=1,\delta_t^+=0,\delta_t^-=0.1$, and the green line represents the result for EWBG with $\delta_h=1,\delta_t^+=0,\delta_t^-=0$. }\label{hhinv}
\end{figure}

The cross section of the Higgs boson pair production depends on all the three
anomalous couplings $\delta_h$, $\delta_t^+$, and $\delta_t^-$, which are all
related to the energy cutoff $\Lambda$.
By convoluting the partonic cross
section in Eq.~(\ref{psigma}) with the parton distribution function $G(x,\mu_f)_{g/P}$, the differential cross section
at hadron level becomes
\begin{equation}
\sigma(PP \to hh)=\int dx_a dx_b G(x_a,\mu_f)_{g/P} G(x_b,\mu_f)_{g/P}   \hat{\sigma}(gg\to hh).
\end{equation}
The numerical results of the normalized invariant mass distributions for the Higgs boson pair are shown in
Fig.~\ref{hhinv}, which is very different from the SM case for the sample points $\delta_h=1,\delta_t^+=0,\delta_t^-=0.1$ and
$\delta_h=1,\delta_t^+=0,\delta_t^-=0$.
In the SM, it only has one peak  located at about $400$ GeV.
But in this EW baryogenesis scenario, it has two peaks, one peak located at about $260$ GeV,
and the other one located at about $420$ GeV.
This character of the invariant mass distribution of the Higgs boson pair can be
used to test this EW baryogenesis scenario induced by the dimension-6 operators.
Due to the difficulties to suppress the backgrounds at the hadron colliders,
it will be difficult to  completely pin down these anomalous couplings at the 14~TeV LHC, even with $3000~\mathrm{fb}^{-1}$ integrated luminosity.
Exploiting  the boosted tricks  may  help to
increase the ability to extract the anomalous coupling.
More precise constraints may come from the future CEPC experiments.

\subsection{Collider phenomenology at the CEPC}
Further, we discuss how to precisely test this EW phase transition and EW baryogenesis scenario  at the CEPC, where the measurements are of great precision.
\subsubsection{Anomalous trilinear Higgs coupling}
An interesting approach is to test these new interactions at next-leading order (NLO) in the associated $hZ$ production \cite{McCullough:2013rea,Englert:2013tya}. The advantage of this method is that the measurement of the $hZ$ production cross section will be precise as $\mathcal{O}(0.1 \%)\sim \mathcal{O}(1\%)$ at future $e^+e^-$ colliders \cite{Baer:2013cma,Azzi:2014jwa}. Although this test for new Higgs self interaction is indirect and may suffer from some theoretical uncertainties \cite{McCullough:2013rea}, it is still a good guide for the future $e^+e^-$ collider to search for the shape of the Higgs potential and the mechanism of EW baryogenesis.

Firstly, we give the  NLO $e^+e^-\rightarrow hZ$ cross section in the SM.
Here, the NLO $e^+e^-\rightarrow hZ$ cross section is calculated in the on-mass-shell renormalization scheme \cite{Denner:1991kt} and includes both radiative corrections from weak and electromagnetic processes as
\begin{equation}
\sigma_\mathrm{NLO} =
\sigma_\mathrm{Born}+\sigma_\mathrm{weak}+\sigma_\mathrm{EM},
\end{equation}
where $\sigma_\mathrm{Born}$ is the Born cross section at tree level, $\sigma_\mathrm{weak}$ is the weak correction, and $\sigma_\mathrm{EM}$ includes contributions from the virtual photon correction to the $eeZ$ vertex and real initial state radiations. We calculate $\sigma_\mathrm{EM}$ in the phase space slicing approach as \cite{Bohm:1993qx,Denner:2003iy}
\begin{equation}
\sigma_\mathrm{EM} =
\sigma_\mathrm{vir}+\sigma_\mathrm{soft}+\sigma_\mathrm{coll}+\sigma_\mathrm{finite},
\end{equation}
where $\sigma_\mathrm{vir}$ is the virtual correction, $\sigma_\mathrm{soft}$ describes the contribution of the soft photon emissions with the energy $E_{\gamma}<\Delta E$, $\sigma_\mathrm{coll}$ contains the contribution of photons collinear to the beam line with $E_{\gamma}>\Delta E$ and $\sin \theta_{\gamma}  < \sin(\Delta \theta)$, and $\sigma_\mathrm{finite}$ includes the remaining correction from the hard photons with $E_{\gamma}>\Delta E$ and $\sin \theta_{\gamma}  > \sin(\Delta \theta)$. $\sigma_\mathrm{NLO}$ would not depend on the two cut parameters $\Delta E$ and $\Delta \theta$, which are introduced to divide the phase space. Here we do not consider the contribution from real photons beyond $\mathcal{O}(\alpha)$, which may be important near the threshold region with $\sqrt{s} \sim m_h+m_Z$~\cite{Denner:2003iy}.

\begin{figure}[!htbp]
\centering
\includegraphics[width=.42\textwidth]{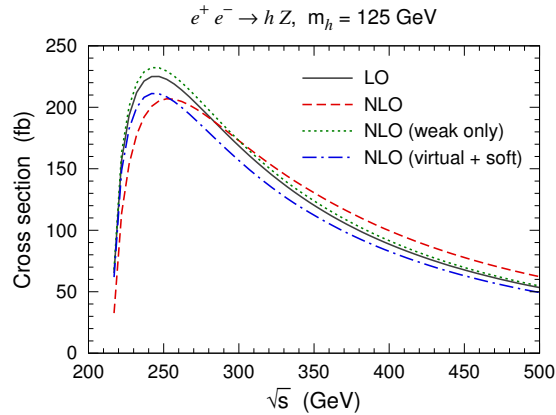}
\caption{The cross section of $e^+e^- \rightarrow h Z$ in the SM. Four lines denote cross sections at LO (solid), NLO (dashed), NLO without the electromagnetic corrections to the $eeZ$ vertex (dot), and NLO with the real corrections only from the soft photons (dot-dashed), respectively. In the last case, we require the photon energy $E_{\gamma}<0.1 \sqrt{s}$, as in Refs.~\cite{Englert:2013tya,Denner:1992bc}.}
\label{fig:Xsec}
\end{figure}
We use the packages \texttt{FeynArts}~\cite{Hahn:2000kx}, \texttt{FormCalc}~\cite{Hahn:1998yk}, and \texttt{LoopTools}~\cite{vanOldenborgh:1990yc} to calculate NLO corrections. Fig.~\ref{fig:Xsec} shows the NLO $e^+e^-\rightarrow hZ$ cross section $\sigma_{hZ}$ in the SM. Detailed investigations can be found in Refs.~\cite{Fleischer:1982af,Denner:1992bc}. It can be seen that the full NLO correction at $\sqrt{s}=250$ GeV reduce $\sigma_{hZ}$ by a factor of $\sim 0.1$.

\begin{figure}[!htbp]
\centering
\includegraphics[width=.42\textwidth]{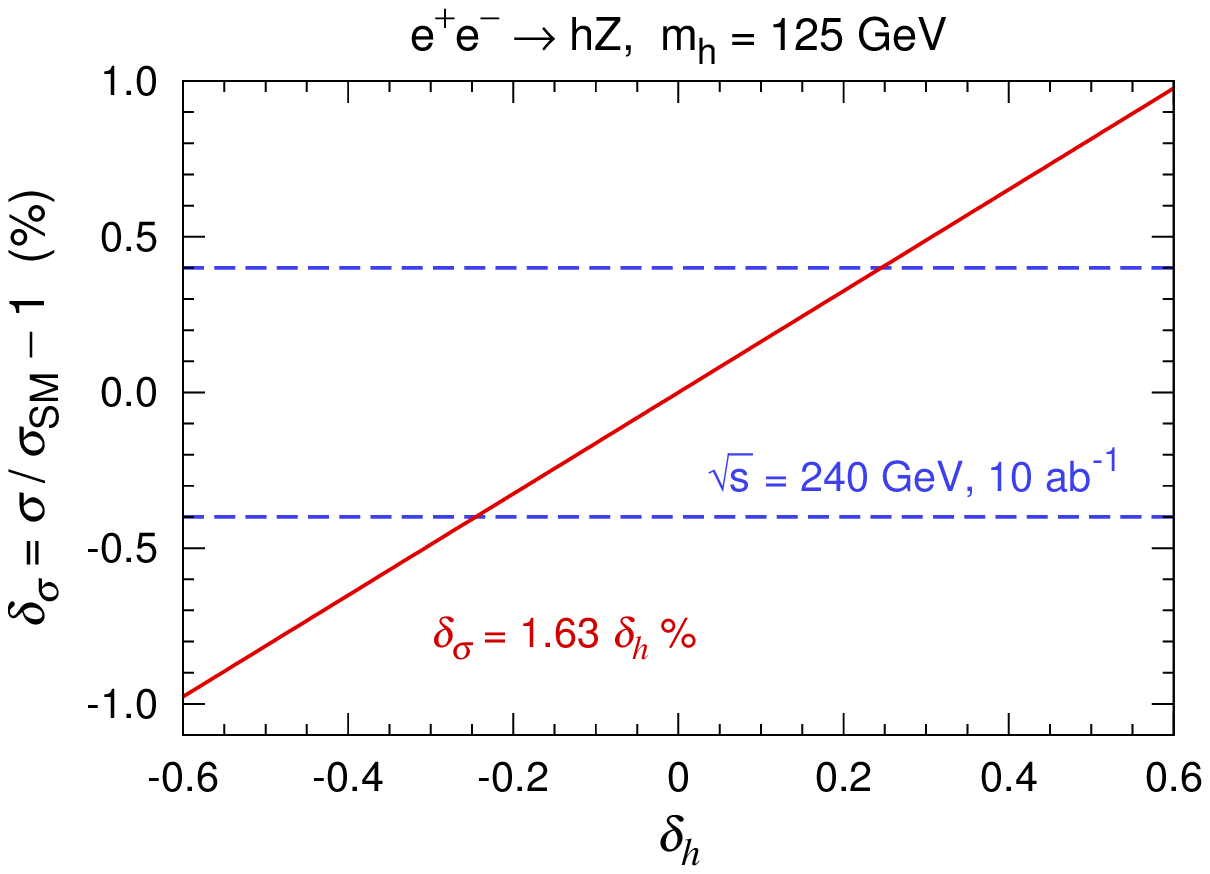}
\caption{The modification of the $e^+e^-\to hZ$ cross section as a function of the anomalous trilinear coupling $\delta_h$ (solid line) at the CEPC with $\sqrt{s}=240$~GeV. The dashed lines denote the sensitivity to $\delta_{\sigma}$ for an integrated luminosity of $10~\mathrm{ab}^{-1}$.}
\label{fig:delh_dXsec_rts240}
\end{figure}

In general, the $hZZ$ coupling and $\sigma_{hZ}$ could be modified by the anomalous Higgs trilinear coupling, anomalous
top quark Yukawa coupling, and even other additional operators in new physics. The generic discussions can be found in Ref.~\cite{McCullough:2013rea}.
The deviation of $\sigma_{hZ}$ can be defined as
\begin{equation}
\delta_{\sigma}=\frac{\sigma_{hZ}}{\sigma_{hZ}^\mathrm{SM}}-1.
\end{equation}
At the CEPC with $\sqrt{s}=240~\GeV$, the contribution to $\delta_{\sigma}$ given by the anomalous Higgs trilinear coupling is much larger than that
given by the anomalous top quark Yukawa coupling~\cite{Shen:2015pha}.
The UV divergences related to $\delta_h$ from different diagrams are automatically canceled.
Assuming $\delta_t^\pm=0$, $\delta_{\sigma}$ is given as a function of $\delta_h$ in Fig. \ref{fig:delh_dXsec_rts240}. We can see that $\delta_{\sigma}$ is approximately proportional to $\delta_h$ as $\delta_{\sigma}\simeq 1.6 \delta_h \%$
at $\sqrt{s}=240~\GeV$. For the future CEPC with an integrated luminosity of $10~\mathrm{ab}^{-1}$, the precision of $\sigma_{zh}$ can be $0.4\%$~\cite{Gomez-Ceballos:2013zzn}. Therefore, it is possible to test $|\delta_h| \sim 25\%$. In our EW baryogenesis scenario with sufficient SFOPT, $\delta_h \in (0.6,1.5)$, which is well within the precision of CEPC.

\subsubsection{Anomalous top Yukawa coupling}

\begin{figure}[!htbp]
\subfigure[\label{fig:vertextth}]
{\includegraphics[width=.25\textwidth]{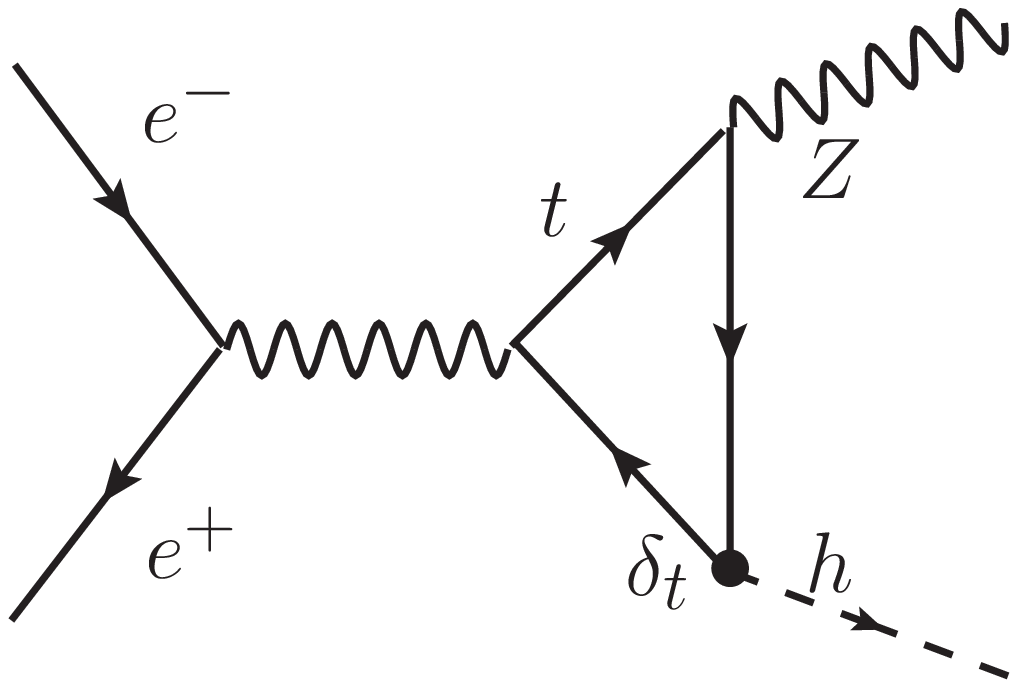}}
\qquad\qquad
\subfigure[\label{fig:selfenergytth}]
{\includegraphics[width=.2\textwidth]{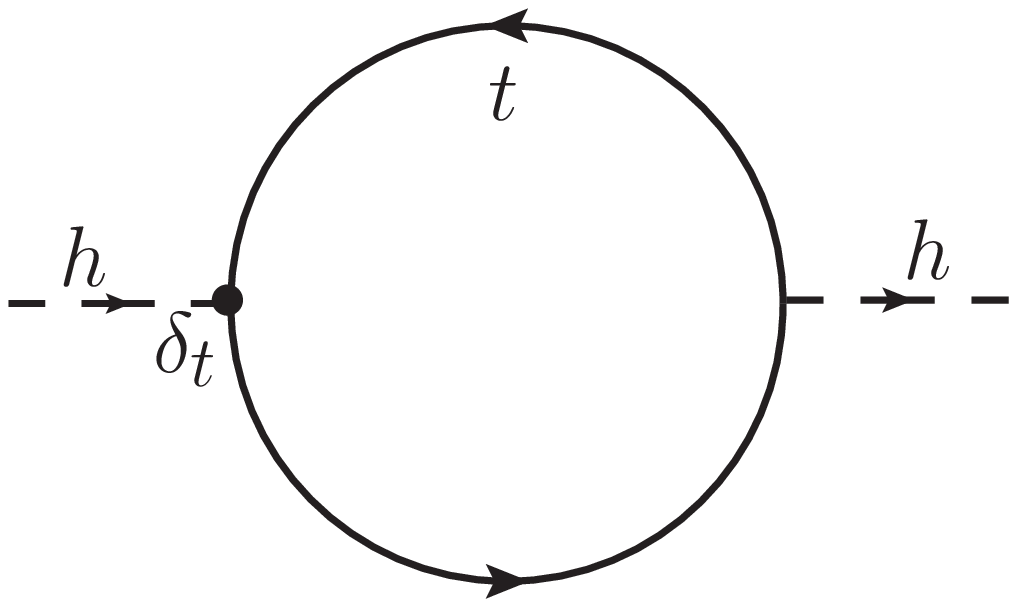}
\includegraphics[width=.2\textwidth]{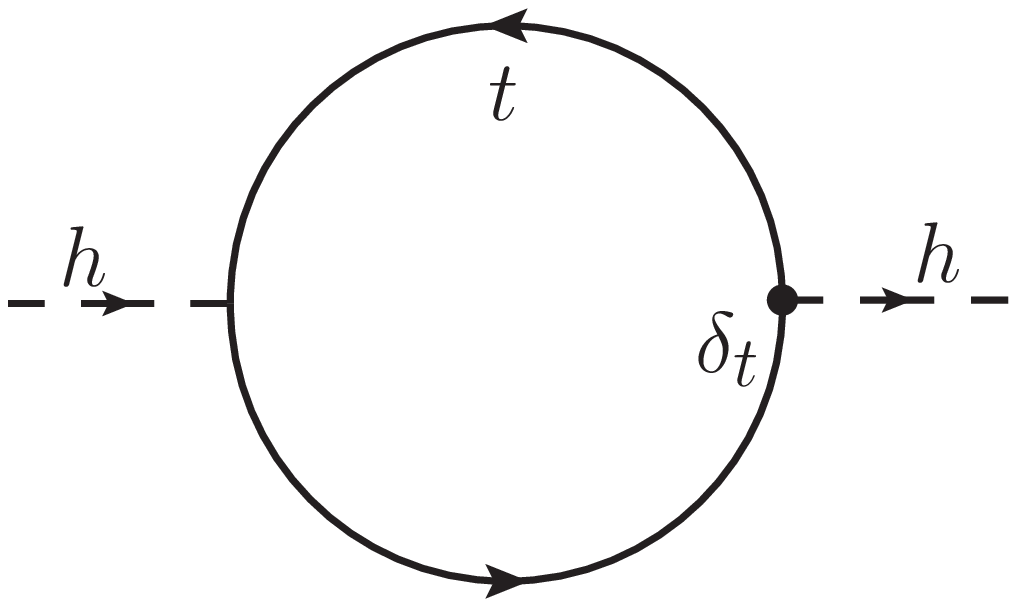}}
\caption{Feynman diagrams describing NLO corrections to $e^+e^-\to hZ$ from the anomalous top quark Yukawa interaction, including the corrections to the $hZ\gamma$ and $hZZ$ vertices (a) and to the Higgs wave function (b).}
\label{fig:diagrammtth}
\end{figure}

In this subsection, we investigate the effect of the anomalous top quark Yukawa coupling.
In general, the scattering amplitude describing interactions between the Higgs and gauge bosons can be parameterized by~\cite{Inoue:2014nva}
\begin{equation}
A(hVV)=\frac{1}{v}(g_0 m_V^2 \epsilon_1 \epsilon_2+g^+ f_{\mu\nu} f^{\mu\nu}+g^-f_{\mu\nu} , \tilde{f}^{\mu\nu}),
\label{effectivehvv}
\end{equation}
where $f^{\mu\nu}=\epsilon^\mu q^\nu-\epsilon^\nu q^\mu$ is the field strength with the polarization vector $\epsilon$ and momentum $q$, and $\tilde{f}^{\mu\nu}=\frac{1}{2} \epsilon^{\mu\nu\alpha\beta}f_{\alpha\beta}$ is the dual field strength tensor. The anomalous top quark Yukawa  couplings enter the $hZZ$ and $hZ\gamma$ vertices and the Higgs wavefunction counter term as shown in Fig.~\ref{fig:diagrammtth}, and contribute to the $g^+$ and $g^-$ terms in Eq.~(\ref{effectivehvv}). The NLO correction comes from the interference between the $g_0$ term at tree level and the $g^+$ and $g^-$ terms at one loop level.

\begin{figure}[!htbp]
\centering
\includegraphics[width=.42\textwidth]{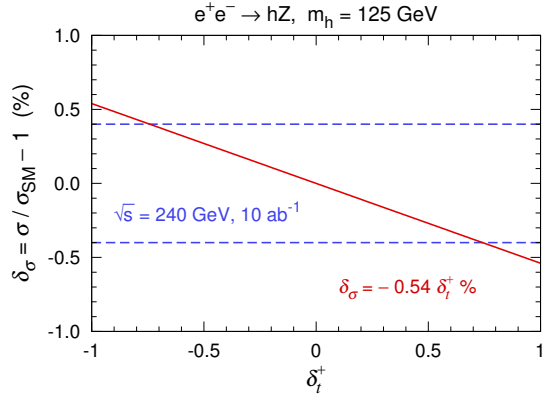}
\caption{The modification of the $e^+e^-\to hZ$ cross section as a function of the anomalous CP-even top quark Yukawa coupling $\delta_t^+$ (solid line) at the CEPC with $\sqrt{s}=240$ GeV. The dashed lines denote the sensitivity to $\delta_{\sigma}$ for an integrated luminosity of $10$ ab$^{-1}$.}
\label{fig:dtp_dXsec_rts240}
\end{figure}

We find that the interference term between the tree level amplitude and the amplitude contributed by the CP-odd top quark Yukawa coupling $\delta^-_t$ vanishes. Therefore this CP-odd coupling does not affect $\sigma_{hz}$ at NLO. In principle, it will contribute to $\sigma_{hz}$ at NNLO with a suppression factor of $\sim (\alpha/4\pi)^2$, but this contribution would be very small and beyond the CEPC sensitivity.
Nonetheless, the anomalous CP-even top quark Yukawa coupling $\delta^+_t$ can alter $\sigma_{hz}$ at NLO.
The vertex corrections introduce a divergence $\propto\delta^+_t$,
which is canceled by the divergence $\propto\delta^+_t$ from the Higgs wave function counter terms in Fig.~\ref{fig:selfenergytth}.
We treat the counter terms that are proportional to $(\delta^+_t)^2$ as higher order contributions and do not include them in the calculation.

\begin{figure}[!htbp]
\centering
\includegraphics[width=.42\textwidth]{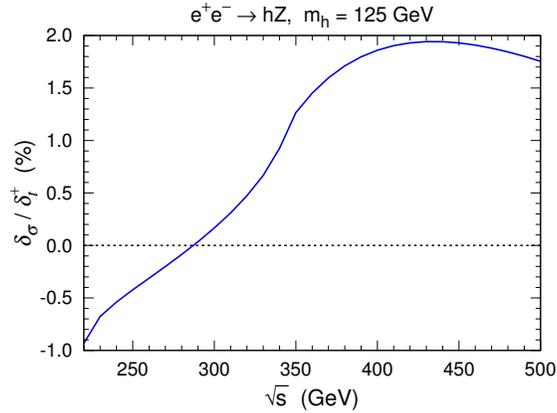}
\caption{$\delta_{\sigma}/\delta_t^+$ as a function of $\sqrt{s}$ for the $e^+e^-\to hZ$ process.}
\label{fig:rts_dXsec_dtp}
\end{figure}

\begin{figure}[!htbp]
\centering
\includegraphics[width=.45\textwidth]{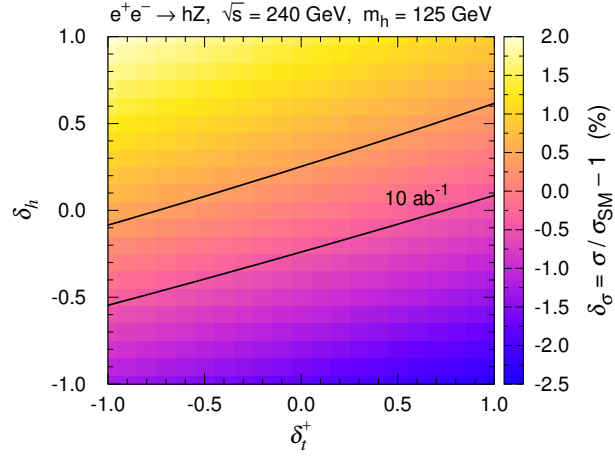}
\caption{The modification of the $e^+e^-\to hZ$ cross section in the $\delta_t^+$-$\delta_h$ plane at the CEPC with $\sqrt{s}=240$ GeV. The sold lines denote the sensitivity to $\delta_{\sigma}$ for an integrated luminosity of $10$ ab$^{-1}$.}
\label{fig:dtp-delh_dXsec_rts240}
\end{figure}

In Fig.~\ref{fig:dtp_dXsec_rts240}, we show the cross section deviation $\delta_\sigma$ as a function of $\delta_t^+$ at $\sqrt{s}=240~\GeV$, assuming $\delta_h=0$.
We find that $\delta_{\sigma} \simeq -0.54 \delta_t^+ \%$.
If $\sigma_{hz}$ can be determined to an accuracy of $0.4\%$, the anomalous coupling would be limited to be $|\delta_t^+|\lesssim 74 \%$. We also give the energy dependence of $\delta_{\sigma}/\delta^+_t$ in Fig.~\ref{fig:rts_dXsec_dtp}. We can see that $\delta_{\sigma}/\delta^+_t$ increases with $\sqrt{s}$. At a collider with $\sqrt{s}=500$ GeV, such as ILC, $\delta_{\sigma}$ $\simeq 1.8 \delta_t^+ \%$. Then it is possible to determine $|\delta_t^+|$ down to $\sim 56 \%$ with a $\sigma_{hz}$ accuracy of $1 \%$ in this indirect approach.
If we also let $\delta_h$ free, $\delta_\sigma$ at the CEPC would vary in the $\delta_t^+$-$\delta_h$ plane as demonstrated in Fig.~\ref{fig:dtp-delh_dXsec_rts240}.
The anomalous couplings would be constrained within the two black solid lines if a data set of $10~\mathrm{ab}^{-1}$ is collected.

\begin{figure}[!htbp]
\centering
\subfigure[\label{fig:ratiog}]
{\includegraphics[width=.45\textwidth]{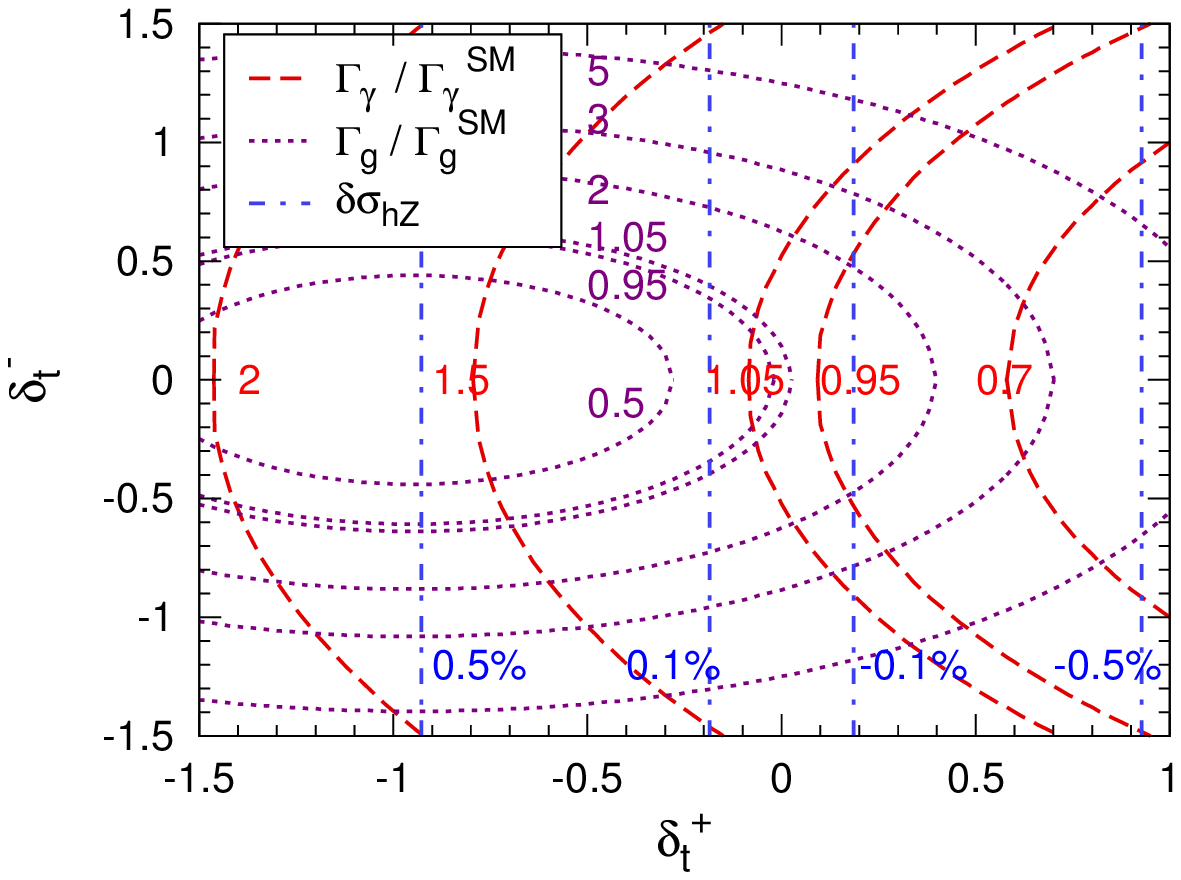}}
\subfigure[\label{fig:ratior}]
{\includegraphics[width=.45\textwidth]{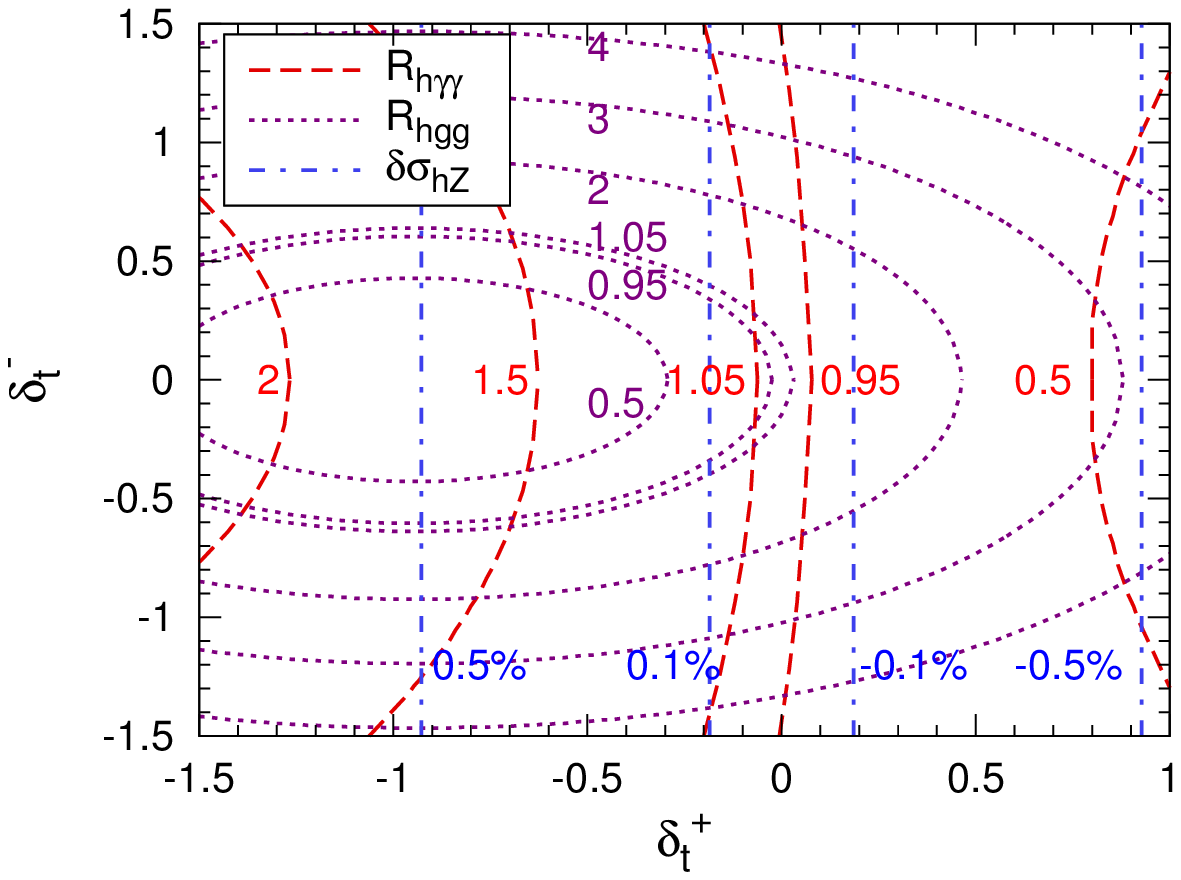}}
\caption{(a) Contours of $\delta_{\sigma}$ (dot-dashed), $\Gamma_{hgg}/\Gamma_{hgg}^\mathrm{SM}$ (dotted), and $\Gamma_{h\gamma\gamma}/\Gamma_{h\gamma\gamma}^\mathrm{SM}$ (dashed) in the $\delta^+_t$-$\delta^-_t$ plane.
(b) Contours of $\delta_{\sigma}$ (dot-dashed), $R_{hgg}$ (dotted), and $R_{h\gamma\gamma}$ (dashed) in the $\delta^+_t$-$\delta^-_t$ plane.}
\label{fig:delpa}
\end{figure}

The anomalous top quark Yukawa coupling would also modify the Higgs couplings to $gg$ and $\gamma\gamma$.  Therefore, the precise measurements of the Higgs partial decay widths could set limits on $\delta^+_t$ and $\delta^-_t$. The deviation of a specified Higgs decay width $\Gamma_{hXX}$  can be described by the signal strength
\begin{equation}
R_{hXX} =\frac{\sigma_{hZ}\cdot \mathrm{Br}(h\rightarrow XX)}{[\sigma_{hZ} \cdot \mathrm{Br}(h\rightarrow XX)]^\mathrm{SM}}
=\frac{\sigma_{hZ}}{\sigma_{hZ}^\mathrm{SM}} \frac{\Gamma_{hXX}}{\Gamma_{hXX}^\mathrm{SM}}\frac{\Gamma_\mathrm{tot}^\mathrm{SM}}{\Gamma_\mathrm{tot}},
\end{equation}
where $\sigma_h$ is the Higgs production cross section, and $\Gamma_\mathrm{tot}$ is the total Higgs decay width.
At the CEPC with an integrated luminosity of $5~\mathrm{ab}^{-1}$, it is possible to determine $R_{hgg}$ and $R_{h\gamma\gamma}$ up to precisions of $\sim 1.6\%$ and $\sim 9\%$, respectively~\cite{CEPC-SPPCStudyGroup:2015csa}. In Fig.~\ref{fig:delpa}, we show $\Gamma_{hgg}/\Gamma_{hgg}^\mathrm{SM}$, $\Gamma_{h\gamma\gamma}/\Gamma_{h\gamma\gamma}^\mathrm{SM}$, $R_{hgg}$, $R_{h\gamma\gamma}$, and $\delta_{\sigma}$ in the $\delta^+_t$-$\delta^-_t$ plane, assuming $\delta_h=0$.
We can see that Higgs decay measurements with an accuracy of $5\%$ will set stringent limits on the anomalous top quark Yukawa couplings up to $\mathcal{O}(10\%)$.

\section{Renormalizable models}\label{sec:ren}

The dimension-6 operators in Eq.~(\ref{dim6}) can be induced from certain renormalizable extensions of the SM. In this section we shall demonstrate a class of realistic models with vector-like quarks and a triplet Higgs.

\subsection{The models with an $[SU(2)_L^{}]$-triplet Higgs without hypercharge}

\begin{figure}
  \centering
  \includegraphics[width=.35\textwidth]{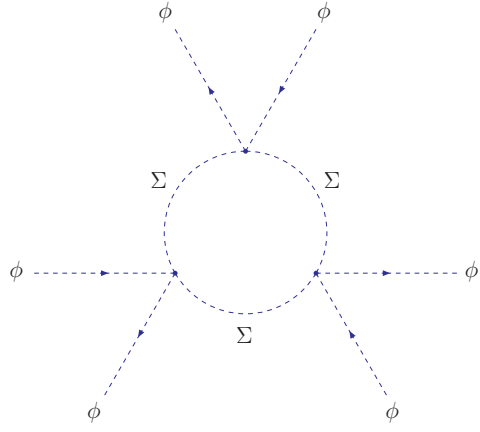}\\
  \caption{One-loop diagram mediated by the Higgs triplet $\Sigma$ without hypercharge for generating the dimension-6 coupling of the Higgs doublet $\phi$.}\label{potential1}
\end{figure}

We first consider the models with an $[SU(2)_L^{}]$-triplet Higgs scalar with a zero hypercharge,
\begin{eqnarray}
\begin{array}{l}\Sigma(1,3,0)\end{array}\!\!&=&\left[\begin{array}{cc}\frac{1}{\sqrt{2}}\sigma^{0}_{}
&\sigma^{+}_{}\\
[2mm]
\sigma^{-}_{}&-\frac{1}{\sqrt{2}}\sigma^{0}_{}\end{array}\right]=\Sigma^\dagger_{}\,.
\end{eqnarray}
The Lagrangian involving the Higgs triplet $\Sigma$ should be
\begin{eqnarray}
\label{tl1}
\delta\mathcal{L}&=&-\xi^{}_\Sigma(\phi^T_{}i\tau^{}_2\Sigma\tilde{\phi}+\textrm{H.c.})
+\frac{1}{2}\textrm{Tr}[(D^\mu_{}\Sigma)^\dagger_{}D_\mu^{}\Sigma]\nonumber\\
&&-\frac{1}{2}M_{\Sigma}^2\textrm{Tr}(\Sigma^2_{})-\frac{1}{4}\zeta_\Sigma^{}[\textrm{Tr}(\Sigma^2_{})]^2_{}
-\frac{1}{2}\kappa_\Sigma^{}\phi^\dagger_{}\phi\textrm{Tr}(\Sigma^{2}_{})\,,
\nonumber\\
&&
\end{eqnarray}
where the covariant derivative is given by
\begin{eqnarray}
D^{}_\mu \Sigma&=&\partial^{}_\mu \Sigma -ig\left[\frac{\tau_a^{}}{2}W^a_\mu,\Sigma\right]\,.
\end{eqnarray}
Before the EW symmetry breaking, the Higgs triplet $\Sigma$ can mediate a one-loop diagram as shown in Fig.~\ref{potential1} to induce the dimension-6 coupling of the SM Higgs doublet $\phi$,
\begin{eqnarray}
\label{loop1}
\mathcal{L}&\supset&-
\frac{1}{\Lambda_{\Sigma}^2}(\phi^\dagger_{}\phi)^3_{}~~\textrm{with}~~
\Lambda_\Sigma^{}=\frac{4\sqrt{2}\pi}{\kappa_\Sigma^{}\sqrt{\kappa_\Sigma^{}}}M_\Sigma\,.
\end{eqnarray}
For a proper choice of the parameters $M_\Sigma^{}$ and $\kappa_\Sigma$, we can fulfill the requirements of the cutoff scale $\Lambda$ in the previous discussion. For example, we can take
\begin{eqnarray}
\label{par1}
M_{\Sigma}^{}=500\,\textrm{GeV}~~\textrm{and}~~\kappa_\Sigma^{}=5~~\Rightarrow ~~\Lambda_\Sigma^{}\simeq 795\,\textrm{GeV}\,.
\end{eqnarray}

\begin{figure}
  \centering
  \includegraphics[width=.5\textwidth]{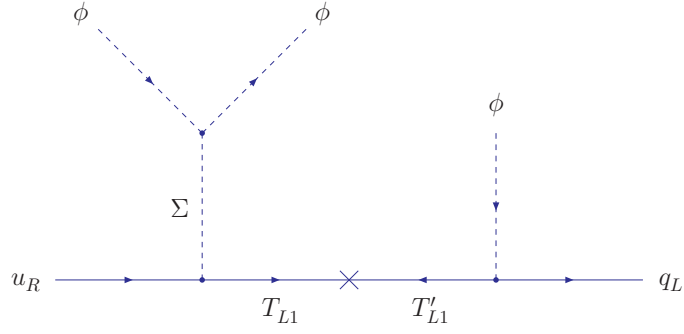}\\
  \caption{\label{yukawa1}
Tree-level diagram mediated by the Higgs triplet $\Sigma$ and the vector-like quark $T1=T^{}_{L1}+T'^{c}_{L1}$ for generating the dimension-6 couplings of the Higgs doublet $\phi$ to the left-handed quarks $q^{}_L$ and the right-handed up-type quarks $u^{}_R$. }
\end{figure}

In order to generate the first term in Eq.~(\ref{dim6}), we further introduce two $[SU(2)_L^{}]$-triplet quarks with the hypercharge $\pm 2/3$,
\begin{eqnarray}
\begin{array}{l}T^{}_{L1}(3,3,+\frac{2}{3})\end{array}\!\!&=&\left[\begin{array}{cc}\frac{1}{\sqrt{2}}T^{+\frac{2}{3}}_{L1}
&T^{+\frac{5}{3}}_{L1}\\
[2mm]
T^{-\frac{1}{3}}_{L1}&-\frac{1}{\sqrt{2}}T^{+\frac{2}{3}}_{L1}\end{array}\right]\,,\nonumber\\
\begin{array}{l}T'^{}_{L1}(3,3,-\frac{2}{3})\end{array}\!\!&=&\left[\begin{array}{cc}\frac{1}{\sqrt{2}}T'^{-\frac{2}{3}}_{L1}
&T'^{-\frac{5}{3}}_{L1}\\
[2mm]
T'^{+\frac{1}{3}}_{L1}&-\frac{1}{\sqrt{2}}T'^{-\frac{2}{3}}_{L1}\end{array}\right]\,,
\end{eqnarray}
which have the Yukawa and mass terms as below,
\begin{eqnarray}
\label{tq1}
\delta\mathcal{L}&\supset&-f'^{}_{1i}\bar{q}^c_{Li}i\tau^{}_2T'^{}_{L1} \tilde{\phi}-f^{}_{1j}\textrm{Tr}(\overline{T}^{}_{L1} \Sigma)u^{}_{Rj}+\textrm{H.c.}
\nonumber\\
&&-M^{}_{T1}[\textrm{Tr}( i\tau^{}_2\overline{T}'^{c}_{L1} i\tau^{}_2 T^{}_{L1})+\textrm{H.c.}]\,.
\end{eqnarray}
We then can obtain the vector-like quark,
\begin{eqnarray}
T1=T^{}_{L1}+T'^{c}_{L1}\,.
\end{eqnarray}
The vector-like quark $T1$ together with the Higgs triplet $\Sigma$ can mediate the dimension-6 couplings of the Higgs doublet $\phi$ to the left-handed quarks $q^{}_L$ and the right-handed up-type quarks $u^{}_R$ at tree level,
\begin{eqnarray}
\label{treet1}
\mathcal{L}&\supset&-\frac{f'^\ast_{1i}f^{}_{1j}\xi_\Sigma^{}}{M_{T1}^{}M_\Sigma^2}\bar{q}^{}_{Li}\tilde\phi u^{}_{Rj}\phi^\dagger_{}\phi+\textrm{H.c.}\,.
\end{eqnarray}
The relevant diagram is shown in Fig.~\ref{yukawa1}.

\begin{figure}
  \centering
  \includegraphics[width=.5\textwidth]{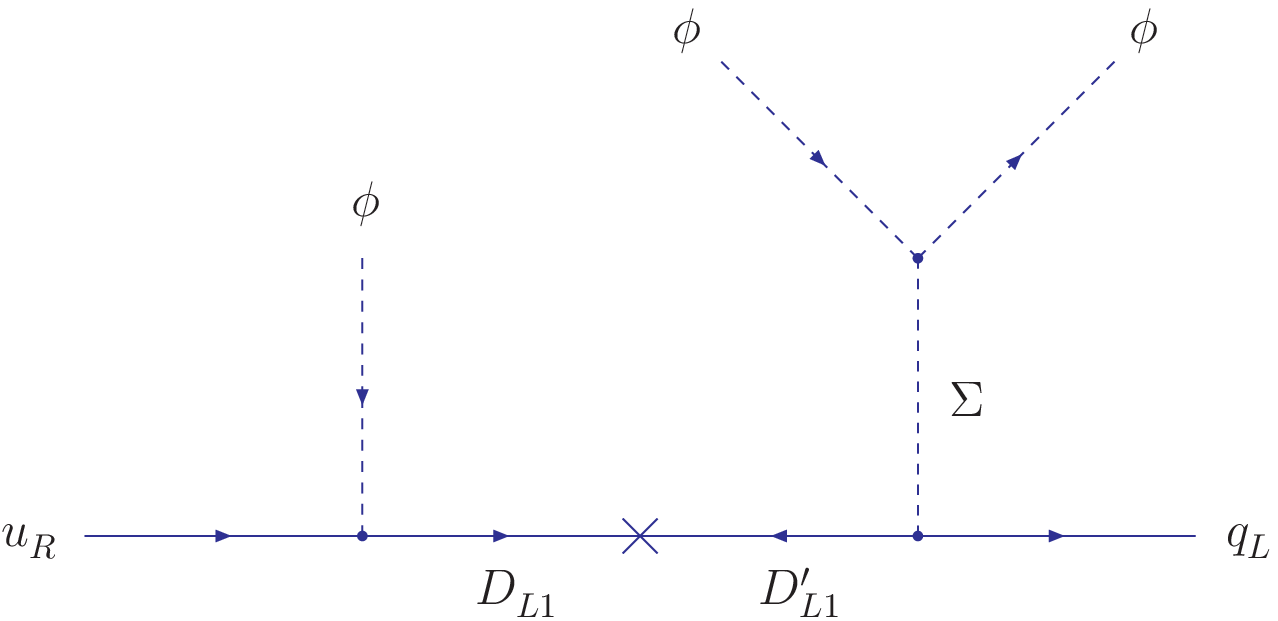}\\
  \caption{\label{yukawa2}
Tree-level diagram mediated by the Higgs triplet $\Sigma$ and the vector-like quark $D1=D^{}_{L1}+D'^{c}_{L1}$ for generating the dimension-6 couplings of the Higgs doublet $\phi$ to the left-handed quarks $q^{}_L$ and the right-handed up-type quarks $u^{}_R$. }
\end{figure}

Alternatively, the first term in Eq.~(\ref{dim6}) can be induced by introducing two $[SU(2)_L^{}]$-doublet quarks with the hypercharge $\pm 1/6$,
\begin{eqnarray}
\begin{array}{l}D^{}_{L1}(3,2,+\frac{1}{6})\end{array}\!\!&=&\left[\begin{array}{c}D^{+\frac{2}{3}}_{L1}
\\
[2mm]
D^{-\frac{1}{3}}_{L1}\end{array}\right]\,,\nonumber\\
\begin{array}{l}D'^{}_{L1}(3,2,-\frac{1}{6})\end{array}\!\!&=&\left[\begin{array}{c}D'^{+\frac{1}{3}}_{L1}
\\
[2mm]
D'^{-\frac{2}{3}}_{L1}\end{array}\right]\,,
\end{eqnarray}
which have the following Yukawa and mass terms,
\begin{eqnarray}
\label{dq1}
\delta\mathcal{L}&\supset&-f'^{}_{1i}\bar{q}^c_{Li}i\tau^{}_2\Sigma D'^{}_{L1} -f^{}_{1j}\overline{D}^{}_{L1} \phi u^{}_{Rj}+\textrm{H.c.}
\nonumber\\
&&-M^{}_{D1}(\overline{D}'^{c}_{L1} i\tau^{}_2 D^{}_{L1}+\textrm{H.c.})\,.
\end{eqnarray}
As shown in Fig.~\ref{yukawa2}, by integrating out the vector-like quark,
\begin{eqnarray}
D1=D^{}_{L1}+D'^{c}_{L1}\,,
\end{eqnarray}
and the Higgs triplet $\Sigma$, we can obtain the expected dimension-6 operator,
\begin{eqnarray}
\label{treed1}
\mathcal{L}&\supset&-\frac{f'^\ast_{1i}f^{}_{1j}\xi_\Sigma^{}}{M_{D1}^{}M_\Sigma^2}\bar{q}^{}_{Li}\phi u^{}_{Rj}\phi^\dagger_{}\phi+\textrm{H.c.}\,.
\end{eqnarray}

Note after the SM Higgs doublet $\phi$ develops its {\rm vev},
\begin{eqnarray}
\langle\phi\rangle=\left[\begin{array}{c}\langle\phi^{0}_{}\rangle
\\
[2mm]
0\end{array}\right]\,,
\end{eqnarray}
to drive the EW symmetry breaking, the Higgs triplet $\Sigma$ will acquire an induced {\rm vev},
\begin{eqnarray}
\langle\Sigma\rangle&=&\left[\begin{array}{cc}\frac{1}{\sqrt{2}}\langle\sigma^{0}_{}\rangle
&0\\
[2mm]
0&-\frac{1}{\sqrt{2}}\langle\sigma^{0}_{}\rangle\end{array}\right]~~\textrm{with}\nonumber\\
&&\langle\sigma^{0}_{}\rangle\simeq -\frac{\sqrt{2}\xi_\Sigma^{}\langle\phi\rangle^2_{}}{\overline{M}^{2}_\Sigma}\,,~~
\overline{M}^{2}_\Sigma=M^{2}_\Sigma+\kappa_\Sigma^{} \langle\phi\rangle^2_{}\,.\quad\quad
\end{eqnarray}
Since the {\rm vev} $\langle\Sigma\rangle$ is constrained by the $\rho$ parameter,
\begin{eqnarray}
\rho=1+\frac{2\langle\Sigma\rangle^2_{}}{\langle\phi\rangle^2_{}}\,,
\end{eqnarray}
it should be convenient to rewrite Eqs.~(\ref{treet1}) and (\ref{treed1}) by
\begin{eqnarray}
\label{eff1}
\mathcal{L}&\supset&
\frac{f'^\ast_{1i}f^{}_{1j}(M^2_\Sigma+\kappa^{}_\Sigma\langle\phi\rangle^2_{})\langle\Sigma\rangle}
{\sqrt{2}M_{F1}^{}M^2_\Sigma\langle\phi\rangle^2_{}}\bar{q}^{}_{Li}\phi u^{}_{Rj}\phi^\dagger_{}\phi+\textrm{H.c.}\nonumber\\
&&\textrm{with}~~M_{F1}^{}=M_{T1}^{}~\textrm{or}~M_{D1}^{}\,.
\end{eqnarray}
Comparing with the above operators in Eq.~(\ref{dim6}), we can determine
\begin{eqnarray}
x^{ij}_{u}=-\frac{16\sqrt{2}\pi^2_{}f'^{\ast}_{1i}f^{}_{1j}(M^2_\Sigma+\kappa^{}_\Sigma\langle\phi\rangle^2_{})
\langle\Sigma\rangle}{\kappa_\Sigma^3 M_{F1}^{}\langle\phi\rangle^2_{}}\,.
\end{eqnarray}
With the parameter choice (\ref{par1}), the CP violation for the EW baryogenesis can be satisfied,
\begin{eqnarray}
x^{33}_{u}\simeq 0.74i\,,
\end{eqnarray}
by inputting
\begin{eqnarray}
\!\!\!\!\!\!\!\!&&f'^{}_{13}=if^{}_{13}=\sqrt{4\pi}\,,~f'^{}_{11}=f'^{}_{12}=f^{}_{11}=f^{}_{12}=0\,,\nonumber\\
\!\!\!\!\!\!\!\!&&M_{F1}^{}=800\,\textrm{GeV}\,,~\langle\sigma^{0}_{}\rangle=2\,\textrm{GeV}\,,~\langle\phi\rangle=174\,\textrm{GeV}\,.\quad\quad
\end{eqnarray}

\subsection{The models with an $[SU(2)_L^{}]$-triplet Higgs with hypercharge}

We now consider the models with an $[SU(2)_L^{}]$-triplet Higgs scalar with a nonzero hypercharge,
\begin{eqnarray}
\begin{array}{l}\Delta(1,3,-1)\end{array}\!\!&=&\left[\begin{array}{cc}\frac{1}{\sqrt{2}}\delta^{-}_{}
&\delta^{0}_{}\\
[2mm]
\delta^{--}_{}&-\frac{1}{\sqrt{2}}\delta^{-}_{}\end{array}\right]\,.
\end{eqnarray}
The scalar interactions of the SM then should be extended by
\begin{eqnarray}
\label{tl2}
\delta\mathcal{L}&=&-\xi^{}_\Delta(\tilde{\phi}^T_{}i\tau^{}_2\Delta\tilde{\phi}+\textrm{H.c.})
+\textrm{Tr}[(D^\mu_{}\Delta)^\dagger_{}D_\mu^{}\Delta]\nonumber\\
&&-M_{\Delta}^2\textrm{Tr}(\Delta^\dagger_{}\Delta)-\zeta_\Delta^{}[\textrm{Tr}(\Delta^\dagger_{}\Delta]^2_{}
\nonumber\\
&&
-\kappa_\Delta^{}\phi^\dagger_{}\phi\textrm{Tr}(\Delta^{\dagger}_{}\Delta)\,,
\end{eqnarray}
with the covariant derivative,
\begin{eqnarray}
D^{}_\mu \Delta&=&\partial^{}_\mu \Delta -ig\left[\frac{\tau_a^{}}{2}W^a_\mu,\Delta\right]+ig'B^{}_\mu \Delta\,.
\end{eqnarray}
Like the Higgs triplet without hypercharge $\Sigma$, the Higgs triplet with hypercharge $\Delta$ can also mediate a one-loop diagram to induce the expected dimension-6 interaction of the SM Higgs doublet $\phi$,
\begin{eqnarray}
\mathcal{L}&\supset&-
\frac{1}{\Lambda^{}_\Delta}(\phi^\dagger_{}\phi)^3_{}~~\textrm{with}~~
\Lambda_\Delta^{}=\frac{4\pi}{\kappa_\Delta^{}\sqrt{\kappa_\Delta^{}}}M_\Delta\,.
\end{eqnarray}
Actually, we can easily obtain
\begin{eqnarray}
\label{par2}
\Lambda_\Delta^{}\simeq 785\,\textrm{GeV}~~\textrm{for}~~M_\Delta=500\,\textrm{GeV}\,,~\kappa_\Delta^{}=4\,.
\end{eqnarray}

We also need two $[SU(2)^{}_L]$-triplet quarks with the hypercharge $\pm 1/3$,
\begin{eqnarray}
\begin{array}{l}T^{}_{L2}(3,3,-\frac{1}{3})\end{array}\!\!&=&\left[\begin{array}{cc}\frac{1}{\sqrt{2}}T^{-\frac{1}{3}}_{L2}
&T^{+\frac{2}{3}}_{L2}\\
[2mm]
T^{-\frac{4}{3}}_{L2}&-\frac{1}{\sqrt{2}}T^{-\frac{1}{3}}_{L2}\end{array}\right]\,,\nonumber\\
\begin{array}{l}T'^{}_{L2}(3,3,+\frac{1}{3})\end{array}\!\!&=&\left[\begin{array}{cc}\frac{1}{\sqrt{2}}T'^{+\frac{1}{3}}_{L2}
&T'^{-\frac{2}{3}}_{L2}\\
[2mm]
T'^{+\frac{4}{3}}_{L2}&-\frac{1}{\sqrt{2}}T'^{-\frac{1}{3}}_{L2}\end{array}\right]\,,
\end{eqnarray}
which have the Yukawa and mass terms,
\begin{eqnarray}
\label{tq2}
\delta\mathcal{L}&\supset&-f'^{}_{2i}\bar{q}^c_{Li}i\tau^{}_2T'^{}_{L2} \phi-f^{}_{2j}\textrm{Tr}(\overline{T}^{}_{L2} \Delta)u^{}_{Rj}+\textrm{H.c.}
\nonumber\\
&&-M^{}_{T2}[\textrm{Tr}( i\tau^{}_2\overline{T}'^{c}_{L2} i\tau^{}_2 T^{}_{L2})+\textrm{H.c.}]\,,
\end{eqnarray}
or two $[SU(2)^{}_L]$-doublet quarks with the hypercharge $\pm\frac{7}{6}$,
\begin{eqnarray}
\begin{array}{l}D^{}_{L2}(3,3,+\frac{7}{6})\end{array}\!\!&=&\left[\begin{array}{c}D^{+\frac{5}{3}}_{L2}
\\
[2mm]
D^{-\frac{2}{3}}_{L2}\end{array}\right]\,,\nonumber\\
\begin{array}{l}D'^{}_{L2}(3,3,-\frac{7}{6})\end{array}\!\!&=&\left[\begin{array}{c}D'^{-\frac{2}{3}}_{L2}
\\
[2mm]
D'^{+\frac{5}{3}}_{L2}\end{array}\right]\,,
\end{eqnarray}
which have the Yukawa and mass terms,
\begin{eqnarray}
\label{dq2}
\delta\mathcal{L}&\supset&-f'^{}_{2i}\bar{q}^c_{Li}i\tau^{}_2\Delta^\dagger_{} D'^{}_{L2} -f^{}_{2j}\overline{D}^{}_{L2} \tilde{\phi} u^{}_{Rj}+\textrm{H.c.}
\nonumber\\
&&-M^{}_{D2}(\overline{D}'^{c}_{L2} i\tau^{}_2 D^{}_{L2}+\textrm{H.c.})\,.
\end{eqnarray}
By integrating out the Higgs triplet $\Delta$ and the vector-like quark,
\begin{eqnarray}
T2=T^{}_{L2}+T'^{c}_{L2}~~\textrm{or}~~D2=D^{}_{L2}+D'^{c}_{L2}\,,
\end{eqnarray}
we can obtain the following dimension-6 operator,
\begin{eqnarray}
\delta\mathcal{L}&\supset&-x_{u}^{ij}\frac{\phi^\dagger_{}\phi}{\Lambda^2_{\Delta}}\bar{q}^{}_{Li}\phi u^{}_{Rj}+\textrm{H.c.}~~\textrm{with}\nonumber\\
&&x^{ij}_{u}=-\frac{16\pi^2_{}f'^{\ast}_{1i}f^{}_{1j}(M^2_\Delta+\kappa^{}_\Delta\langle\phi\rangle^2_{})
\langle\Delta\rangle}{\kappa_\Delta^3 M_{F2}^{}\langle\phi\rangle^2_{}}\,,\nonumber\\
&&M_{F2}^{}=M_{T2}^{}~\textrm{or}~M_{D2}^{}\,.
\end{eqnarray}
In the presence of the parameter choice (\ref{par2}), a desirable CP violation can be
\begin{eqnarray}
\!\!\!\!\!\!\!\!&&x^{33}_{u}=0.95i~~\textrm{for}\nonumber\\
\!\!\!\!\!\!\!\!&&f'^{}_{13}=if^{}_{13}=\sqrt{4\pi}\,,~f'^{}_{11}=f'^{}_{12}=f^{}_{11}=f^{}_{12}=0\,,\nonumber\\
\!\!\!\!\!\!\!\!&&M_{F2}^{}=800\,\textrm{GeV}\,,~\langle\Delta\rangle=2\,\textrm{GeV}\,,~\langle\phi\rangle=174\,\textrm{GeV}\,.
\end{eqnarray}

It should be noted like the {\rm vev} $\langle\Sigma\rangle$, the {\rm vev} $\langle\Delta\rangle$ which is given by
\begin{eqnarray}
\langle\Delta\rangle&=&\left[\begin{array}{cc}0
&\langle\delta^0_{}\rangle\\
[2mm]
0&0\end{array}\right]~~\textrm{with}\nonumber\\
&&\langle\delta^{0}_{}\rangle\simeq -\frac{\xi_\Delta^{}\langle\phi\rangle^2_{}}{\overline{M}^{2}_\Delta}\,,~~
\overline{M}^{2}_\Delta=M^{2}_\Delta+\kappa_\Delta^{} \langle\phi\rangle^2_{}\,,\quad\quad
\end{eqnarray}
is also constrained by the $\rho$ parameter,
\begin{eqnarray}
\rho=1+\frac{2\langle\Delta\rangle^2_{}}{\langle\phi\rangle^2_{}+2\langle\Delta\rangle^2_{}}\,.
\end{eqnarray}
Furthermore, the Higgs triplet $\Delta$ can have the Yukawa couplings with the SM left-handed leptons,
\begin{eqnarray}
\delta\mathcal{L}\supset - y\bar{l}^{c}_{L}i\tau^{}_2 \Delta^\dagger_{} l^{}_L +\textrm{H.c.}~~\textrm{with}~~\begin{array}{l} l_L^{}(1,2,-\frac{1}{2})\end{array}=\left[\begin{array}{c}\nu^{}_L\\
[2mm]
e^{}_L\end{array}\right]\,.\nonumber\\
&&
\end{eqnarray}
Due to the small neutrino masses, the Yukawa couplings $f$ should be tiny when the {\rm vev} $\langle\Delta\rangle$ is expected at the GeV scale. To avoid this fine tuning, we can arrange the Higgs triplet $\Delta$ for a zero lepton number to forbid the above Yukawa couplings if the lepton number is exactly conserved or is only softly broken.

\section{Conclusion}\label{sec:sum}

Unravelling the true shape of the Higgs potential, the type of the phase transition and the  EW baryogenesis
is the central  issue after the discovery of the Higgs boson, and is a major goal for the future CEPC.
We have investigated the simple extension of the Higgs sector
by introducing the dimensions-6 operators $-x_{u}^{ij}\frac{\phi^\dagger_{}\phi}{\Lambda^2_{}}\bar{q}^{}_{Li}\tilde{\phi} u^{}_{Rj}+\textrm{H.c.}-\frac{\kappa}{\Lambda^2_{}}(\phi^\dagger_{}\phi)^3$ using the EFT and
discussed how to test this scenario at colliders.
We have found that the $\frac{\kappa}{\Lambda^2_{}}(\phi^\dagger_{}\phi)^3$ operator could  provide another possible Higgs potential with the same
Higgs mass and {\rm vev},
easily realize the SFOPT, and modify  the
trilinear Higgs boson coupling obviously.
The sizable CP violation source can be supplied by the
operator $-x_{u}^{ij}\frac{\phi^\dagger_{}\phi}{\Lambda^2_{}}\bar{q}^{}_{Li}\tilde{\phi} u^{}_{Rj}+\textrm{H.c.}$
, which can induce the anomalous top quark Yukawa coupling.
Both the anomalous trilinear Higgs boson coupling and the anomalous top quark  Yukawa coupling can make contributions
to the Higgs boson pair production at the LHC, and the analytical expressions  and the simple numerical results are given in this paper.
The invariant mass distribution of  Higgs boson pair at the LHC
is expected to be different from the SM prediction to achieve SFOPT and EW baryogenesis.
However, due to the precision limit of the LHC, the CEPC is needed to
precisely test this scenario, and
the anomalous coupling can be tested indirectly from the precise measurements of the $Zh$ production at the CEPC.
Besides the investigation using the EFT to get the model independent predictions, concrete renormalizable
models are built, and they can give  the concerned dimension-6 operators.
The study will help us to understand the nature of EW symmetry breaking and the origin of BAU from the
current LHC experiments and the future CEPC experiments.

\begin{acknowledgments}
We thank Xiao-Jun Bi for early cooperation and useful discussion.
F.P.H. and X.Z. are supported by the NSFC under grants Nos. 11121092, 11033005, 11375220
and also by the CAS pilotB program.
P.H.G. is supported by the Shanghai Jiao Tong University under Grant No. WF220407201, the Recruitment Program for Young Professionals under Grant No. 15Z127060004, the Shanghai Laboratory for Particle Physics and Cosmology under Grant No. 11DZ2260700.
P.F.Y. and Z.H.Y. are supported by the NSFC under grants No. 11475189.
Z.H.Y. is also supported by the Australian Research Council.
\end{acknowledgments}

\section{appendix}
The scalar integrals are defined as:
\begin{eqnarray}
C_{ij}&=&\int \frac{d^4q}{i\pi^2}~\frac{1}
{(q^2-m_t^2)( (q+p_i)^2-m_t^2)
( (q+p_i+p_j)^2-m_t^2)}  \,, \\
D_{ijk}&=&\int \frac{d^4q}{i\pi^2} \frac{1}
{(q^2-m_t^2)( (q+p_i)^2-m_t^2)
( (q+p_i+p_j)^2-m_t^2)( (q+p_i+p_j+p_k)^2-m_t^2)} \,.
\end{eqnarray}
The analytic expressions can be found in Ref.~\cite{Ellis:2007qk}.

\end{document}